\documentclass{elsart}
\usepackage{natbib}
\usepackage{amssymb}
\usepackage{graphicx}

\newcommand\lax{\>\vcenter{\hbox{$<$\hskip-.75em\lower1.0ex\hbox{$\sim$}}}\>}
\newcommand\uax{\>\vcenter{\hbox{$>$\hskip-.75em\lower1.0ex\hbox{$\sim$}}}\>}

\begin{document} 

\begin{frontmatter}

\title{Development of the Gaseous Antiparticle Spectrometer for Space-based 
                                               Antimatter Detection}   

\author{Charles J. Hailey\thanksref{label1}}, \author{W.W. Craig\thanksref{label2}}, \author{F.A. Harrison\thanksref{label3}}, \author{J. Hong\thanksref{label4}}, \author{K. Mori\thanksref{label1}},
\author{J. Koglin\thanksref{label1}}, \author{H.T. Yu\thanksref{label1}} and \author{K.P. Ziock\thanksref{label2}}

\address[label1]{ Columbia Astrophysics Laboratory, Columbia University, 550 W. 120th St., NY,
NY 10027} 
\address[label2]{University of California, Lawrence Livermore National Laboratory, P.O. Box
808, Livermore, CA 94550 }
\address[label3]{California Institute of Technology, 1200 E. California Blvd., Pasadena, CA
91125} 
\address[label4]{ Harvard College Observatory, Harvard University, B-418,
Cambridge, MA 02138} 

\maketitle 

\begin{abstract}

We report progress in developing a novel antimatter detection scheme. The
gaseous antiparticle spectrometer (GAPS) identifies antimatter through the
characteristic X-rays emitted by antimatter when it forms exotic atoms in
gases. The approach provides large area and field of view, and excellent
background rejection capability. If the GAPS concept is successfully
demonstrated, then it would be an ideal candidate for space-based, indirect
dark matter searches. GAPS can detect antideuterons produced in neutralino
annihilations. A modest GAPS experiment can detect the neutralino for all
minimal SUSY models in which the neutralino mass is in the $\sim50$--$350$ GeV mass
range. Underground searches, by contrast, are only sensitive to about $1/2$ the
SUSY parameter space in this mass range.   
\end{abstract}

\end{frontmatter}

\section{Introduction} 

A major goal of 21st century physics is to identify particle dark matter. The
leading candidate for this dark matter is the neutralino, a weakly interacting
massive particle arising from supersymmetric extensions of the standard model
\citep{jungman96}.  The neutralino is the lightest supersymmetric partner and
is stable under R-conservation.  

There are a number of experiments underway around the world to detect the
neutralino through the nuclear recoils produced through its scalar and vector
coupling to matter. These experiments are done deep underground. Only upper
limits on the neutralino mass and couplings have been obtained.  Third
generation experiments are planned which would have target masses of $\uax$ 1
metric ton \citep{aprile02}.   

An alternative approach is indirect detection of the neutralino. The
neutralino is a Majorana particle and therefore its own antiparticle. When
neutralinos annihilate in the galactic halo they produce hadronic
showers. These showers contain antibaryons, most notably antiprotons. These
primary antiprotons have been the subject of intensive searches by the
cosmic-ray community. However the difficulty of detecting them is well-known
\citep{simon98, bergstrom98}.  Secondary antiprotons produced in cosmic-ray collisions with hydrogen
and helium in the interstellar medium swamp the neutralino-produced primary
antiproton signal. The problem is exacerbated by solar modulation effects. The
primary antiproton signal can only be detected at $\lax$ 0.1 GeV. Such antiproton
energies are only accessible outside the heliosphere, and deep space probes
carrying antiproton detectors have been proposed \citep{wells99}.   

\begin{figure}
\begin{center} 
\includegraphics*[width=8cm]{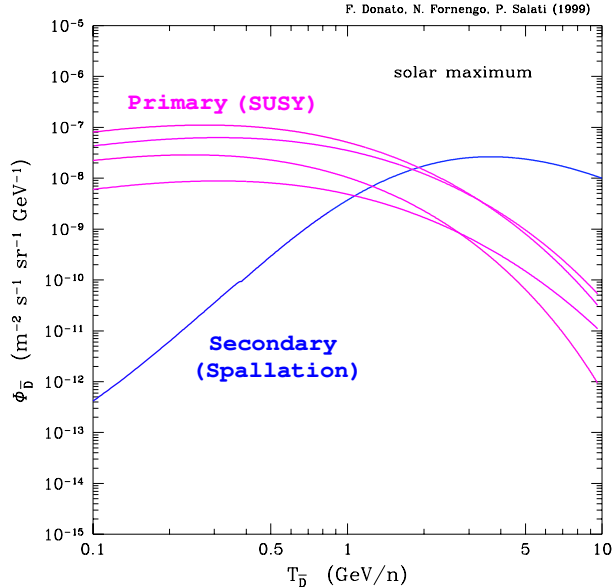}
\end{center} 
\caption{Solar modulated flux of primary and secondary antideuterons. The
primary antideuteron flux curves correspond to four cases with different
neutralino masses ($M_{\chi}$), gaugino-higgsino mixing ratios ($P_g$) and
neutralino relic abundances ($\Omega_\chi h^2$) \citep{donato00}.}
\end{figure} 

The promise of indirect detection has changed radically in the last couple of
years. Theoretical calculations predict a measurable flux of primary
antideuterons formed in neutralino annihilation \citep{donato00, donato01}. A primary antideuteron
search is vastly superior to an antiproton search for the neutralino. The
primary antideuteron flux increases markedly at lower energies. And the
secondary antideuteron flux, produced in cosmic-ray hydrogen collisions, has a
much higher kinematic threshold for production. This combined with the steep
falloff of cosmic-ray protons with increasing energy leads to a sharply
falling secondary antideuteron flux at low energies. This is illustrated in
fig. 1. Therefore a search for primary antideuterons is feasible in low
earth orbit. 

We are developing a novel detector concept called the 
gaseous antiparticle spectrometer (GAPS) which has been specifically
invented for antideuteron searches.  A description of how GAPS
would be employed in a space-based antideuteron search, as well as detailed
physics motivation for such an indirect search are presented in Mori et
al. 2002.  

\section{Principle of GAPS operation} 

\begin{figure}
\begin{center} 
\includegraphics*[width=9cm]{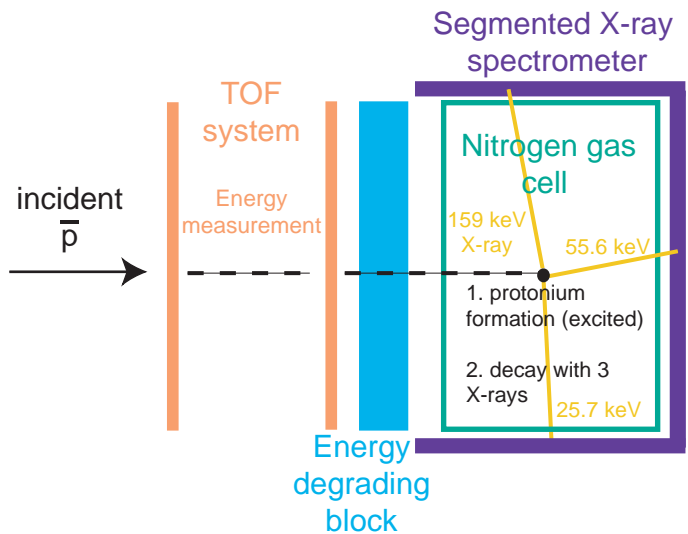}
\end{center} 
\caption{The principle of GAPS for antiparticle detection illustrated with an
antiproton. }
\end{figure} 

A simplified schematic of the GAPS detector is shown in fig 2.  The
antiparticle passes through the time of flight (TOF) system (which measures
its velocity) and is slowed down by $dE/dx$ losses in a degrader.  
The thickness of the degrader is tuned to ensure the antiparticle beam is
slowed down and stopped in the gas chamber. An exotic atom is formed with
probability of order unity. The exotic atom is in a high excitation state, and
deexcites through a process involving both autoionizing transitions and
radiation producing transitions. Through proper selection of gas and gas
pressure the absorption of the antiparticle can be tailored to produce 3-4
well-defined X-ray transitions in the exotic atom decay chain.  The X-rays
have energies in the $\sim25$--250 keV range so that the gas and the
surrounding gas chamber support structure are optically thin to them. These X-rays are
absorbed in an NaI(Tl) X-ray spectrometer (but a solid state detector like CZT
is equally feasible) which surrounds the gas cell. Promptly after the release
of these X-rays the antiparticle annihilates with the nucleus producing a
shower of pions. The coincident signal between the TOF system, the
characteristic decay X-rays and the energy deposition of the pions is an
extremely clean criterion for the presence of an antiparticle. The energies of
the 3 or 4 X-rays uniquely define the type of antiparticle through the Bohr
transition energies. This approach also allows us to suppress background with
rejection factors $\uax 10^{12}$ since the typical space-based particle and gamma-ray
background produces uncorrelated energy depositions.      

\section{Crucial Technical Issues to be addressed in a GAPS prototype}

Most of the atomic physics issues of relevance to GAPS have been elucidated
both theoretically and experimentally by a variety of groups. However
assembling this work into a coherent picture of how GAPS might operate is
challenging and leaves enough questions to warrant laboratory
investigations. Accelerator testing will be crucial for detector optimization
and simulations of ultimate performance. The atomic physics issues have been
comprehensively addressed elsewhere \citep{mori02}.  All the atomic physics
considerations addressed below are equally applicable to either antiproton or
antideuteron detection in GAPS. Simple atomic physics scaling laws allow one
to relate either theoretical or experimental results to another type of
antiparticle.  

The most important issue to be addressed in accelerator testing is the yield
 of X-ray ladder transitions. This yield can be quenched by Stark
 mixing. Shortly after the capture of the antiparticle, the bound electrons in
 the exotic atom are completely ionized through autoionization.   
 The electric field from adjacent atoms distorts the different angular
 momentum quantum states, inducing $\Delta n   
= 0$ transitions. These transitions lead the exotic atom to an $S$-state followed
 by a nuclear annihilation or to an $nS\rightarrow 1S$ radiative transition. The photon
 energy from the $nS\rightarrow 1S$ radiative transition in appropriate GAPS gases is too
 high to be measured by thin NaI(Tl) or CZT detectors as would be employed in
 a real space detector. Therefore it is crucial to ensure that the Stark
 effect is minimized in a GAPS detector. The detailed theory has been worked
 out \citep{mori02} so that the yield of X-rays as a function of gas type and pressure is
 predicted. Laboratory tests at the appropriate gas pressure and gas
 composition must be done to confirm predictions. The Stark effect is the main
 factor providing constraints on GAPS operating pressure. This is crucial
 since the gas pressure is a design driver; the gas pressure defines the
 thickness of the degrader which affects both the efficiency for detecting
 antiparticles (through absorption losses) and the time resolution of the
 detector (through the finite travel time effects in the degrader).  Other
 yield effects include Coulomb deexcitation, by which transition energy is
 transferred to kinetic energy of the exotic atom and abrupt ladder
 terminations, due to annihilation of the antiparticle in the nucleus. This
 effect should be small \citep{reifenrother02} but this needs to be confirmed. 
 Similarly yield calculations may assume the probability of antiparticle capture into angular
momentum states is statistically distributed, or may invoke more sophisticated
distributions \citep{yamazaki02}. The effects of such assumptions can be tested
in laboratory experiments. More detailed considerations can be found in
 \citet{mori02}.    

Currently we are developing a GAPS prototype that will be tested at the KEK accelerator
facility in Japan in spring 2004. The prototype will be exposed to an antiproton beam. 
A layout of the experiment is shown in fig. 3. The plastic scintillators provide
timing to distinguish antiprotons from pions and kaons, which are also present in
the beam. The plastic in front and back of the gas cell allow us to identify
antiprotons which are not absorbed in the gas cell. These antiprotons are both moving
slowly and produce a large energy deposit in the back plastic. The gas cell is 20 cm
in length and about 5 cm in radius. This provides adequate stopping power for the
antiprotons and allows us to capture $>$ 90\% of the antiprotons emerging from the degrader.The
GAPS gas cell consists of a reinforced carbon composite which holds 1-20 atmospheres
of gas.  The walls are thin enough to offer high transmission
to the lowest X-ray energies of interest, $\sim$20 KeV.  There are 8 hexagonal
panels surrounding the gas cell to provide good solid angle coverage.
Each panel has a $2 \times
4$ array of 25 mm diameter, 5 mm thick NaI(Tl) scintillator crystals which are
optically isolated from each other and which detect the X-rays. The number 
of crystal cells was
designed to ensure that the pions produced in the ground state annihilation of
the exotic atom do not have a large probability of being absorbed in a crystal
cell in which an X-ray is absorbed. Each crystal is coupled to a Hamamatsu
RM1924 photomultiplier tube. The crystals can resolve the 3--4 X-ray 
transitions of interest. We will test Nitrogen, Oxygen, Neon and Argon gas in
our experiment.  For instance, typical transitions of interest in Neon gas are
4 to 3 (115.8 keV), 5 to 4 (53.6 keV) and 6 to 5 (29.1 keV) transition.  Results on X-ray yield and overall antiproton loss in the
degrader and detector material will be compared with our Monte-Carlo
predictions.  

\begin{figure}
\begin{center} 
\includegraphics*[width=8.5cm]{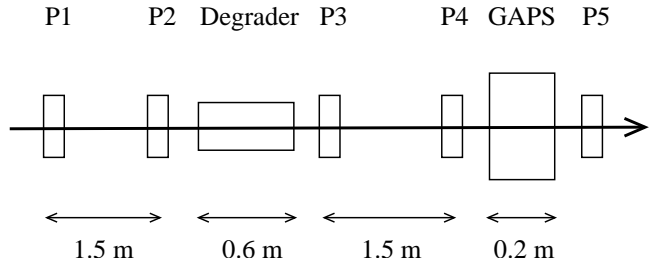}
\end{center} 
\caption{GAPS prototype detector. P = trigger counter (plastic scintillator).}
\end{figure} 

Antiproton testing can serve as an excellent surrogate for antideuteron
performance with regards atomic physics effects. There is only one area where
the antiproton tests do not provide information of direct utility. This
concerns antiparticle losses before the formation of the exotic atom. The
dominant loss mechanism for antiprotons is the well-known direct annihilation
channel.  However antideuterons have additional loss channels which are
familiar from their deuteron analogs \citep{peaslee48}. These include Coulomb
disintegration \citep{gold63} and the Oppenheimer-Phillips \citep{oppenheimer35} process. In the O-P
process the deuteron is torn apart as the antineutron is attracted and the
antiproton repelled from the nucleus. This effect is negligible in the
antideuteron because of the attractive Coulomb force. The Coulomb
disintegration occurs due to photodissociation of the antideuteron in the
virtual photon field it encounters as it passes through matter. We have
calculated this effect by integrating the experimentally known cross-section
for this process along the antideuteron path for all the materials the
antideuteron would encounter in a GAPS detector. The losses due to all of
these nuclear processes are $\lax$ 5--10\%. 

This work was performed with support from the NASA SR\&T program under grant
NAG5-5393.

\end{document}